# An Adaptive QRS Detection Algorithm for Ultra-Long-Term ECG Recordings

John Malik[1], Elsayed Z Soliman[2], Hau-Tieng Wu[134]

[1]Department of Mathematics, Duke University, Durham, NC

[2]Epidemiological Cardiology Research Center (EPICARE), Department of Epidemiology and Prevention, and Department of Medicine, Cardiology Section, Wake Forest School of Medicine, Winston-Salem, NC

[3]Department of Statistical Science, Duke University, Durham, NC

[4]Mathematics Division, National Center for Theoretical Sciences, Taipei, Taiwan

# Abstract

**Background:** Accurate detection of QRS complexes during mobile, ultra-long-term ECG monitoring is challenged by instances of high heart rate, dramatic and persistent changes in signal amplitude, and intermittent deformations in signal quality that arise due to subject motion, background noise, and misplacement of the ECG electrodes. **Purpose**: We propose a revised QRS detection algorithm which addresses the above-mentioned challenges. **Methods and Results:** Our proposed algorithm is based on a state-of-the-art algorithm after applying two key modifications. The first modification is implementing local estimates for the amplitude of the signal. The second modification is a mechanism by which the algorithm becomes adaptive to changes in heart rate. We validated our proposed algorithm against the state-of-the-art algorithm using short-term ECG recordings from eleven annotated databases available at Physionet, as well as four ultra-long-term (14-day) ECG recordings which were visually annotated at a central ECG core laboratory. On the database of ultra-long-term ECG recordings, our proposed algorithm showed a sensitivity of 99.90% and a positive predictive value of 99.73%. Meanwhile, the state-of-the-art QRS detection algorithm achieved a sensitivity of 99.30% and a positive predictive value of 99.68% on the same database. The numerical efficiency of our new algorithm was evident, as a 14-day recording sampled at 200 Hz was analyzed in approximately 157 seconds.

**Conclusions:** We developed a new QRS detection algorithm. The efficiency and accuracy of our algorithm makes it a good fit for mobile health applications, ultra-long-term and pathological ECG recordings, and the batch processing of large ECG databases.

# Introduction

Many computer algorithms have been developed to automatically detect QRS complexes from digital ECG tracings [1]. Some of these algorithms are intended to be used at the bedside to provide real-time monitoring of a patient's heart rate (HR) and cardiac rhythm. Other algorithms are intended to be used in post-processing applications, where batches of ECG signals are analyzed or classified offline. Current algorithms achieve accuracies exceeding 99.9% when subjects demonstrate normal sinus rhythm and when their ECG recordings are short in duration and free of noise and motion artifacts. However, there is still a need for improvement in certain clinical and research situations. For example, in heart rate variability (HRV) analysis, constructing an accurate normal R-to-R interval time series is critical. A QRS detection algorithm with even 99.9% accuracy will fail to detect approximately 750 QRS complexes in a 14-day ECG signal (and generate the same number of erroneous detections), and the HRV analysis result will be deteriorated unless a tedious manual correction is carried out. Needless to say, ECG recordings are rarely free of noise. In such recordings, the task of automatic QRS detection is more challenging, and the performance of most algorithms is far from perfect.

Besides noise, the presence of arrhythmias and intermittent conduction defects (manifested as irregularly timed heart beats) is another major challenge. Morphologically speaking, pathological beats rarely resemble normal sinus beats. Within a given recording, the size and polarity of the QRS complexes may vary significantly. These challenges are more prominent in long-term ECG recordings, where QRS complex detection is confounded by subject motion and the possibility of misplaced ECG electrodes. Long-term recordings often resemble a patchwork of markedly different short recordings (some of which are unreadable), and any algorithm which utilizes global properties of the recording will likely fail. Low computational complexity is required for both real-time analysis and the analysis of ultra-long-term signals. In this manuscript, we propose a revised QRS detection algorithm which addresses the above-mentioned challenges. Specifically, we built and validated an algorithm for automatic QRS detection suitable for the analysis of ultra-long-term ($\geq$14 days) and pathological ECG signals. Our validation process included testing on an annotated database of 14-day ECG recordings and multiple databases of standard ECG recordings.

# Methods

Our proposed algorithm is based on an algorithm proposed by Elgendi [2] with two key modifications. The first modification is the implementation of local estimates for the amplitude of the signal. The second modification is a mechanism by which a parameter in Elgendi's algorithm adapts to changes in heart rate. Below, we provide the technical details of our algorithm. In the appendix, we provide the technical details of Elgendi's original algorithm.

## **Our QRS Detection Algorithm**

Write the raw, single-channel ECG signal as an $n$-dimensional vector $x \in \mathbb{R}^n$, where $n = \lfloor f_s \times T \rfloor$, $f_s$ is the sampling rate of the signal, and $T$ is the duration of the recording in seconds.

Write the $i$-th entry of $x$ as $x(i)$. Begin by applying a 3rd order, bi-directional, Butterworth bandpass filter with cutoff frequencies 8 Hz and 20 Hz. Denote the filtered signal as $y \in \mathbb{R}^n$. Form the vector $z := y \odot y$ by squaring the entries of $y$. Then, apply three moving-average filters to $z$. The first filter has a window size of $W_1$, where $W_1$ is the smallest odd integer greater than or equal to $0.097 \times f_s$. The third filter has a window size of $W_3$, where $W_3$ is the smallest odd integer greater than or equal to $5f_s$. The second filter has a variable window size $W_2 \in \mathbb{R}^n$; we will discuss how to obtain $W_2$ in the next subsection. Compute $v_1 = \texttt{movmean}(z, W_1)$, $\bar{z} = \texttt{movmean}(z, W_3)$, and

$$v_2(i) = \frac{1}{W_2(i)}\left[z\left(i - \frac{W_2(i)-1}{2}\right) + \cdots + z(i) + \cdots + z\left(i + \frac{W_2(i)-1}{2}\right)\right] \quad (1)$$

averaging over the available samples when near the endpoints of the signal $z$. To locally estimate the baseline noise level in the signal, set $\alpha = 0.08 \times \bar{z}$. To detect QRS complexes in the recording $x$, look for sections of the signal $v_1$ which exceed the signal $v_2 + \alpha$ for a duration of at least $W_1$ consecutive samples. The search begins by creating a logical vector

$$L(i) = \begin{cases} 1 & \text{if } v_1(i) > v_2(i) + \alpha(i) \\ 0 & \text{if } v_1(i) \leq v_2(i) + \alpha(i) \end{cases} \quad (2)$$

and applying a moving-sum filter to $L$ with window size $W_1$; $v_3 = \texttt{movsum}(L, W_1)$. Finally, conclude that a QRS complex exists at sample $j$ if there exist positive integers $l_1 \leq j \leq l_2$ such that all of the following conditions hold.

- $v_3(i) = W_1$ for all $l_1 \leq i \leq l_2$
- $v_3(l_1 - 1) \neq W_1$ or $l_1 = 1$
- $v_3(l_2 + 1) \neq W_1$ or $l_2 = n$
- $j = \text{argmax}_{l_1 \leq i \leq l_2} v_1(i)$

The condition $v_3(i) = W_1$ means that in the signal $L$, the window centered at sample $i$ with width $\frac{W_1-1}{2}$ contains only 1's. The first, second, and third conditions encapsulate the search for consecutive sections of $L$ which are positive. The last condition encapsulates the search for the exact location $j$ for the QRS complex (within the previously identified window).

*Adapting to Changes in Heart Rate*

In this subsection, we describe the procedure for calculating the entries of the vector $W_2$. Broadly speaking, we locally estimate the heart rate using the short-time Fourier transform, and we use these local estimates to impute the entries of $W_2$. Choose a discrete window function $h \in \mathbb{R}^{2K+1}$ which satisfies $h(K+1) = 1$. We take the Hann window [3], defined as

$$h(i) = \frac{1}{2}\left[1 - \cos\left(\frac{\pi(i-1)}{K}\right)\right] \qquad (3)$$

In our implementation, we take the window width to be $K = \lfloor 2.5 \times f_s \rfloor$. Set $M = 2f_s$ to be the number of Fourier modes. For each integer (second) $1 \leq t \leq T$ and each integer $m$ satisfying $3 \leq m \leq 25$, evaluate

$$G_t(m) = \sum_{k=1}^{2K+1} [v_1(tf_s + k - K - 1) - \mu_t]h(k)e^{-\frac{2\pi i(k-1)(m-1)}{2M}} \qquad (4)$$

where $v_1(l) := 0$ when $l < 1$ or $l > n$, and

$$\mu_t = \frac{1}{2K+1}\sum_{k=1}^{2K+1} v_1(tf_s + k - K - 1) \qquad (5)$$

is the mean of $v_1$ in the $t$-th window. Note that $G_t(m)$ is the row-$t$, column-$m$ entry of the discretized time-frequency representation. (Due to speed and memory concerns, we compute only a submatrix.) To extract the dominant curve $p$ in the spectrogram, set the time-1 entry to be

$$p(1) = \mathrm{argmax}_{3 \leq m \leq 25}|G_1(m)|^2 \qquad (6)$$

and for each integer $1 < t \leq T$, set the time-$t$ entry to be

$$p(t) = \text{argmax}_{3 \leq m \leq 25} \left[ \frac{|G_t(m)|^2}{\sum_{i=3}^{25} |G_t(i)|^2} - \lambda (m - p(t-1))^2 \right] \quad (7)$$

The constant $\lambda$ controls the smoothness of the curve $p$ by preventing large jumps in the frequency axis between consecutive time points. In our implementation, we take $\lambda = 0.01$. Construct a signal $F \in \mathbb{R}^n$ so that $F(i)$ is an estimate for the current heart rate (in Hz, to the nearest 15 beats-per-minute) by first setting

$$F(tf_s) = \frac{[p(t) - 1] \times f_s}{2M} \quad (8)$$

and then computing the remaining values of $F$ via nearest-neighbor interpolation. Finally, to obtain the desired variable window $W_2$, set (in accordance with Bazett's formula [4])

$$W_2(i) = \left\lfloor \frac{0.611 \times f_s}{\sqrt{F(i)}} \right\rfloor \quad (9)$$

while adding 1 to any entry of $W_2$ that is even.

*Numerical Implementation*

We implemented Elgendi's original algorithm and our new QRS detection algorithm in MATLAB 2019a using the built-in functions. Parameters for our new QRS detection algorithm were either inherited from Elgendi's original algorithm (which was trained on the MIT-BIH Arrhythmia Database) or chosen in an *ad hoc* fashion. In particular, no parameter optimization was performed. The bandpass filter was constructed using the function `butter`. The moving average filters were implemented using the function `movmean`. The moving sum was implemented using the function `movsum`. The variable-window moving average filter was implemented by computing all the necessary fixed-window moving averages and combining the resulting signals appropriately. For the purpose of reproducibility, the MATLAB code is available per request. A real-time implementation of our algorithm would, due to the size and position of the Hann window used in the short-time Fourier transform, provide QRS complex locations after a delay of 2.5 seconds.

## **Validation Databases**

### *Eleven Conventional ECG Databases*

Following Elgendi's procedure [2], we validated our algorithm on eleven annotated databases available at Physionet [5]. The databases are the AF Termination Challenge Database [6], the Fantasia Database [7], the Intracardiac Atrial Fibrillation Database, the MIT-BIH Arrhythmia Database [8], the MIT-BIH Noise Stress Test Database [9], the MIT-BIH Normal Sinus Rhythm Database [10], the MIT-BIH ST Change Database [11], the MIT-BIH Supraventricular Arrhythmia Database [12], the QT Database [13], the St. Petersburg Institute of Cardiological Technics (INCART) 12-lead Arrhythmia Database, and the T-wave Alternans Challenge Database [14]. The MIT-BIH Normal Sinus Rhythm Database has the largest number (1,729,629) of annotated QRS complexes. We used the first available lead in all databases. All databases except the Fantasia Database present recordings with at least two leads (except some in the MIT-BIH ST Change Database), and in the appendix, we report the performance of both algorithms on the second lead (when available). The Normal Sinus Rhythm Database was modified before processing by removing the noise at the end of each recording. Noise removal was done by removing sections that were at least one second after the last QRS complex annotation. Contrary to Elgendi's procedure [2], we did not remove any records from the Fantasia Database. All records and annotations were imported into MATLAB using the WaveForm DataBase (WFDB) Toolbox [15].

*Database of Ultra-Long-Term ECG Recordings*

To the best of our knowledge, despite the increasing prevalence of ultra-long-term ECG monitoring [16, 17], an annotated, publicly available database of ultra-long-term and pathological ECG recordings is lacking. Our new database of single-lead, ultra-long-term ECG recordings comprises four recordings; each recording is approximately two weeks (14 days) in length. The data was recorded using the ZIO® Patch cardiac monitor (iRhythm Technologies, Inc., San Francisco, California, USA) at a sampling rate of 200 Hz. The underlying information of the subjects was unknown to us [18]. Across the four 14-day recordings, we randomly selected 1,200 ten-second segments for manual annotation. These segments underwent manual annotation at the Epidemiological Cardiology Research Center (EPICARE Center, Wake Forest School of Medicine, Winston Salem, NC). To speed up the annotation process, the ECG core laboratory was provided with the estimated QRS complex locations generated by our new QRS detection algorithm. The quality of each ECG segment was also documented as part of the annotation process. After excluding 16 of the 1,200 ECG segments deemed totally unreadable by the ECG core laboratory, 1,184 segments with 15,605 QRS complex annotations remained, all of which were included in our analysis. Of the 15,605 annotated QRS complexes, 1,232 were labeled as ectopic.

## **Performance Evaluation**

Using the twelve annotated ECG databases described above, we compared our new QRS detection algorithm with Elgendi's original algorithm. We used three evaluation metrics: sensitivity (SE), positive predictive value (PPV), and the F1 score (F1). Most QRS detection algorithms in the literature rely on metrics defined in terms of true positives (TP), false positives (FP), and false negatives (FN) [19, 20]. TP is defined as the number of annotations which are "matched" to a predicted QRS complex location, FP is defined as the number of predictions which are not matched to any annotation, and FN is defined as the number of annotations which are not matched to any prediction. We use the standard grace period of 150 ms [21]. SE and PPV are then defined as

$$\text{SE} = 100\% \times \frac{\text{TP}}{\text{TP} + \text{FN}} \quad \text{PPV} = 100\% \times \frac{\text{TP}}{\text{TP} + \text{FP}} \quad (10)$$

The F1 score is the harmonic mean of SE and PPV. All calculations were performed in MATLAB 2019a on an Intel i7-4790K processor. Our beat-by-beat comparison algorithm is an implementation of the process described in [21] and is available per request.

# Results

**Table 1** shows the performance of our new algorithm and Elgendi's original algorithm on twelve annotated ECG databases. Recall that SE reflects the probability that a QRS complex will be detected, and PPV reflects the probability that a predicted QRS complex location actually corresponds to a true QRS complex. The F1 score is the harmonic mean of SE and PPV. The performance of our new QRS detection algorithm is comparable with the performance of Elgendi's original algorithm on the eleven conventional ECG databases available at Physionet. The largest SE value in **Table 1** is 99.99%, which was achieved by our algorithm on the QTDB. This result means that for records in the QTDB, only one beat out of every 10,000 beats was missed by our algorithm. The largest PPV value in **Table 1** is 99.83%, which was also achieved by our algorithm on the QTDB. This result means that out of every 10,000 predictions made by our algorithm, only 17 did not correspond to true QRS complexes. Excluding the IAFDB, the largest difference in SE between our new algorithm and Elgendi's algorithm across the eleven conventional databases is +0.53% on the AFTDB. On the 14-day ECG database, the difference in SE is larger at +0.60%. The significance of this result can be appreciated when we consider that a 14-day ECG recording will have in excess of 1.5 million beats; a decrease in SE of 0.1% means the missed detection of 1,500 QRS complexes. The difference in PPV on the 14-day ECG database is a modest +0.05%. The cost of the improved performance is evidently a doubling in computation time. Nevertheless, the numerical efficiency of our new algorithm is evident, as a 14-day recording sampled at 200 Hz can be analyzed in approximately 157 seconds (just over two-and-a-half minutes). Computation time did not include reading of the electrocardiogram into the workspace, and the reported values in **Table 1** were not weighted according to the sampling rate or length of the records. Note that our reported performance for Elgendi's original algorithm is different than the performance reported by Elgendi [2], and this difference could be the result of different procedures for metric evaluation (a grace period is not specified in [2], nor is a beat-by-beat comparison algorithm), updates to the conventional databases, a different choice of lead, or a different signal for the final QRS demarcation.

## Visualization of the Adaptive Threshold

We show in **Figure 1** a segment of our new database of long-term and pathological recordings wherein Elgendi's algorithm fails due to the high heart rate (approximately 180 bpm in this 7-second extract). In this application of Elgendi's algorithm, the threshold (shown in red) is too high because the fixed window size $W_2$ is large relative to the subject's QT interval length. The signal $v_1$ is shown in black. The predicted QRS complex locations are shown in blue, and one can immediately notice five QRS complexes that were not detected by Elgendi's algorithm. After estimating the heart rate during this section of the recording and applying Bazett's formula to adjust $W_2$, the window size reduces from 611 ms to approximately 353 ms. We show in Figure 2 the same segment of $v_1$ with the new adaptive threshold $v_2 + \alpha$ plotted in red. Notice that the false negative peaks from Figure 1 are now detected in Figure 2.

# Discussion

In this report, we introduced a new QRS detection algorithm which could be a better fit for mobile, ultra-long-term cardiac ECG monitoring due to its ability to adapt to high heart rates, adjust to changing amplitudes, and ignore previous noise and motion artifacts. We validated our algorithm on eleven standard ECG databases and on a new annotated database of ultra-long-term ECG recordings.

## Summary of Previous Work

The algorithm of Pan and Tompkins [19] was one of the first automatic QRS detectors and remains the most highly cited ECG annotation algorithm in the literature. Since then, several QRS detection algorithms have been proposed. Below, we summarize these algorithms to make the novelty of the algorithm that we will propose more apparent.

### *The Algorithm of Pan and Tompkins*

Elements of Pan and Tompkins' algorithm persist in innovations today. The ECG signal is first filtered in such a way that noise is attenuated and QRS complexes are emphasized. Then, QRS complexes are located by finding peaks in the filtered ECG signal. (The value of the filtered ECG signal at any point in time can be thought of as the probability that a QRS complex is located there.) The algorithm of Pan and Tompkins features a complicated thresholding algorithm that attempts to adapt to changes in heart rate by looking at the previous eight detected heart contractions. As described by Elgendi [2], this method is error-compounding because incorrectly detected QRS complexes at the beginning of the signal cause succeeding estimates to also be incorrect.

*Continual Development of QRS Detection Algorithms*

Following Pan and Tompkins [19], several algorithms were proposed which demonstrate higher prediction accuracies. We refer readers to a summary of these works [1], in which the authors describe a plethora of QRS detection algorithms; they further note the lack of evidence for automatic QRS detectors being implemented and tested in clinics. Numerous other surveys have been produced in recent decades [22-27], and we also refer readers with interest to these publications. In this work, based on our own experience, we consider the state-of-the-art QRS detector to be the algorithm proposed by Elgendi [2].

*Elgendi's Algorithm*

Elgendi's QRS detection algorithm [2] was designed to be a numerically efficient alternative to the plethora of recently developed high-accuracy QRS detectors. Such detectors commonly score in excess of 99% sensitivity and positive predictive value on the MIT-BIH Arrhythmia Database [1]. Elgendi's algorithm was validated on the eleven conventional ECG databases considered in this work. Parameters for his algorithm were trained on the MIT-BIH Arrhythmia Database. Since the publication of Elgendi's results, the databases appear to have changed; for example, records in the MIT Normal Sinus Rhythm Database are now around 24 hours long, but he reported that the longest record was 130 minutes in duration. We mention that Elgendi's algorithm does not adapt to changes in heart rate and that it relies on global properties of the recording. In the appendix, we provide a technical description of Elgendi's algorithm, on which our new algorithm is based.

## **Two Key Modifications**

Below, we discuss the two significant differences between our algorithm and Elgendi's algorithm. We also discuss two minor technical differences.

*Local Estimates for the Baseline Noise Level*

We do not want motion artifacts or large measurement noise in one section of the recording to affect QRS detection protocol in another section of the recording. In our algorithm, estimation of the (relative) amplitude is done by the signal $\bar{z}$, and estimation of the baseline noise level is done by the signal $\alpha$. The vector $\alpha$ is "local" in the sense that $\alpha(i)$, the estimated baseline noise level at time $i/f_s$, is calculated using a five-second window surrounding the time of interest. Large increases in amplitude during other parts of the recording have no effect on our QRS detection protocol during the central time period.

*Adapting to Changes in Heart Rate*

As previously discussed, Elgendi's algorithm does not adapt to changes in heart rate. While Elgendi's algorithm performs well on many types of pathological signals, we found that it possesses a particular weakness; when the subject's heart rate is very high, Elgendi's algorithm fails to detect all of the QRS complexes (see **Figure 1**). Our algorithm has been designed to overcome this weakness. Our algorithm succeeds when the parameter $W_2(i)$ is approximately equal to the expected length of the QT interval (or PT interval) at time $i/f_s$. To estimate the expected length of the QT interval, we use the fact that QT interval length is inversely related to instantaneous heart rate [28]. To be specific, our algorithm estimates the instantaneous HR and then uses this estimate to impute $W_2(i)$. Estimating the instantaneous HR is simple when the locations of the QRS complexes are known. However, we must estimate the HR without knowing the locations of the QRS complexes. Hence, we chose to use the short-time Fourier transform to obtain an estimate for the HR. The discrete Fourier transform could be applied directly to the signal $x$ to obtain an estimate for the average HR observed during the recording. However, for long-term ECG signals, the HR may change significantly from time to time.

Consequently, our HR estimate must be made locally. In our implementation of the short-time Fourier transform, we use a window size of five seconds. The maximum argument of the power spectrum of this five-second window is approximately one fifth the number of heart beats occurring during those five seconds.

*Minor Modifications*

On top of the two significant differences discussed above, there are two minor technical differences between our algorithm and Elgendi's algorithm that we should mention. First, our QRS complex detection criterion is slightly different than the one used by Elgendi (see Appendix). In our algorithm, the final location $j$ of the demarcated QRS complex is chosen from a smaller range of samples. Second, we note that in Elgendi's original work, the signal to be used in the definition of $j$ was not specified. Elgendi simply writes,

> "The last stage is finding the maximum absolute value within each block, the R peak"
>
> [2].

We have chosen to use $v_1$ to demarcate the QRS complex because in terms of performance on the annotated MIT-BIH Arrhythmia Database, this setting for Elgendi's algorithm was the best of several options. This choice is further discussed below.

## **QRS Complex Demarcation**

Normal QRS complexes are typically around 80 ms in duration, and any QRS detection algorithm has the additional task of placing a QRS marker at some point in this short interval. Small differences in QRS marker position can have a strong effect on secondary ECG analysis tools such as the R-to-R interval time series. Conventionally, the QRS marker must be placed at the R peak. In individuals free of cardiac arrhythmias, the R wave is easy to identify. However, for some non-sinus beats, there is no standard approach to annotating the Q, R, and S waves.

Hence, when we cannot rule out the possibility of encountering non-sinus beats, we use the maximum of the signal $v_1$ to represent the "center" of the QRS complex, or the instance at which QRS-related spectral energy is the highest. As can be seen in **Figure 1**, the $v_1$ maxima do not always correspond to the highest amplitude peaks in the corresponding QRS complexes. If the subject is known to demonstrate normal sinus rhythm, then an additional programming step can identify the R wave using each predicted QRS complex location provided by our algorithm. (Note that this step has not been applied to the signal in **Figure 1**).

## **Signal Quality**

Sixteen segments (out of 1,200) in the 14-day ECG database were deemed to be unreadable by the ECG core laboratory that annotated the database. Our algorithm made QRS complex predictions in these segments, and these predictions were thrown out before calculating SE and PPV. Similarly, we found that at the end of each record in the NSRDB are sections of noise with zero manual QRS complex annotations; we removed all predictions in these poor-quality ECG segments before calculating our evaluation metrics. Including these segments in the evaluation would lead to an unfairly low PPV score. The remedy for poor signal quality is of course ensuring good subject preparation and correct application of the ECG electrodes. Nevertheless, our QRS complex detection algorithm should be deployed with an accompanying automatic signal quality classifier. One efficient signal quality index which we have found works well relies on the fact that when two QRS complex detection algorithms disagree significantly, the signal is likely of low quality [29]. We would not recommend pairing our new QRS complex detection algorithm with Elgendi's algorithm because of the high degree of similarity. Instead, we recommend an algorithm which employs the first or second derivative of the signal; changing the definition of $z := y \odot y$ in our algorithm to $z := y' \odot y'$ would be sufficient. A suitable signal quality index for ultra-long-term ECG signals will be reported in future work.

## Technical Details

On the technical front, we mention that the adjustment of $W_2$ is done using Bazett's formula because of its common use, despite its flaws. To further improve the algorithm, we will consider other nonlinear relationships between heart rate and QT interval length [28] or adopt an alternative QT interval correction method in light of the current ECG recommendations [30].

### *Grace Period*

In the literature, the "grace period" for marking the agreement between an annotation and a prediction is usually a generous 150 ms or is not stated, but a minority of publications choose 100 ms [31-33] or 50 ms [22]. It is not clear whether these publications use a matching approach to performance evaluation as in the ANSI/AAMI standard [21]. It is important to note that without the use of a standardized algorithm for matching predicted QRS complex locations to reference annotations, performance metrics are easily confounded. For example, an algorithm which makes two predictions for every annotation would achieve perfect scores.

## Study Limitations

Our study has limitations. First, due to limited resources, we randomly selected for annotation only 1,200 ten-second segments from the four 14-day ECG signals. While we do see some degree of heterogeneity among the segments randomly selected for annotation, a larger database with more subjects and more annotations is needed to further confirm the performance and generalizability of our algorithm. To speed up the manual annotation process, initial estimates for the QRS complex locations (made by our algorithm) were provided to the ECG core laboratory. While this approach saved time and labor, the reference labels could be biased to our proposed algorithm. This limitation has been mitigated by visually analyzing the labeled ECG segments for any possible bias, but developing reference annotations from scratch is the approach which avoids this issue. The algorithm was implemented in MATLAB and is available per request. The authors suspect that a direct implementation of the filter in C or C++ would speed up the algorithm, which is critical for mobile health applications.

## Conclusions

We introduce an adaptive QRS detection algorithm for single-channel ECG which achieved a sensitivity of 99.90% and a positive predictive value of 99.73% on our newly annotated database of ultra-long-term and pathological ECG signals. State-of-the-art performance was also demonstrated on eleven publicly available ECG databases. Features of our algorithm include adaptations to current heart rate (estimated via the short-time Fourier transform) and local estimates for the amplitude of the signal. The numerical efficiency of our algorithm makes it a good fit for mobile health applications and for the batch processing of large databases of long-term ECG signals. The algorithm is robust to noise (as demonstrated by its performance on the MIT-BIH Noise Stress Test Database), requires zero human input, and showed strong performance across a variety of sampling and bit rates. The signal is also not required to be normalized or to be in units of mV.

## Acknowledgements

The authors thank the staff and faculty at EPICARE; especially Dr. Yabing Li and Mr. Charles Campbell for helping with the annotation and programming of the 14-day ECG dataset.

Disclosures: None

ADAPTIVE QRS DETECTION 20

Table 1: Evaluation of Algorithm Performance on Twelve Annotated ECG Databases

| Database | Number of Beats | Algorithm | SE (%) | PPV (%) | F1 (%) | Compute Time (ms) |
|---|---|---|---|---|---|---|
| AFTDB | 7,590 | This work | 99.67 | 99.04 | 99.35 | 6.77 |
|  |  | Elgendi | 99.14 | 99.08 | 99.11 | 4.21 |
| FANTASIADB | 285,311 | This work | 99.89 | 99.73 | 99.81 | 904.63 |
|  |  | Elgendi | 99.90 | 99.76 | 99.83 | 505.02 |
| IAFDB | 7,637 | This work | 86.96 | 78.38 | 82.45 | 100.60 |
|  |  | Elgendi | 85.98 | 79.80 | 82.77 | 59.76 |
| MITDB | 109,494 | This work | 99.92 | 99.62 | 99.77 | 320.99 |
|  |  | Elgendi | 99.88 | 99.79 | 99.83 | 187.64 |
| NSTDB | 25,590 | This work | 96.51 | 82.82 | 89.14 | 327.45 |
|  |  | Elgendi | 96.44 | 83.47 | 89.49 | 186.40 |
| NSRDB | 1,729,629 | This work | 99.91 | 99.32 | 99.61 | 5,662.07 |
|  |  | Elgendi | 99.92 | 99.32 | 99.62 | 2,923.07 |
| STDB | 76,175 | This work | 99.92 | 99.36 | 99.64 | 312.26 |
|  |  | Elgendi | 99.94 | 99.37 | 99.65 | 179.90 |
| SVDB | 184,583 | This work | 99.77 | 99.65 | 99.71 | 132.99 |
|  |  | Elgendi | 99.69 | 99.70 | 99.70 | 71.11 |
| QTDB | 86,995 | This work | 99.99 | 99.83 | 99.91 | 117.62 |
|  |  | Elgendi | 99.98 | 99.79 | 99.88 | 68.11 |
| INCARTDB | 175,906 | This work | 98.11 | 95.08 | 96.57 | 237.66 |
|  |  | Elgendi | 98.24 | 95.44 | 96.82 | 136.14 |
| TWADB | 18,991 | This work | 98.50 | 94.94 | 96.68 | 30.26 |
|  |  | Elgendi | 98.34 | 94.92 | 96.60 | 18.77 |
| 14DAYDB | 15,605 | This work | 99.90 | 99.73 | 99.82 | 156,508.03 |
|  |  | Elgendi | 99.30 | 99.68 | 99.49 | 77,762.80 |

*Note*: The databases are the AF Termination Challenge Database (AFTDB), the Fantasia Database (FANTASIADB), the Intracardiac Atrial Fibrillation Database (IAFDB), the MIT-BIH Arrhythmia Database (MITDB), the MIT-BIH Noise Stress Test Database (NSTDB), the MIT-BIH Normal Sinus Rhythm Database (NSRDB), the MIT-BIH ST Change Database (STDB), the MIT-BIH Supraventricular Arrhythmia Database (SVDB), the QT Database (QTDB), the St. Petersburg Institute of Cardiological Technics 12-lead Arrhythmia Database (INCARTDB), the T-wave Alternans Challenge Database (TWADB), and our ultra-long-term (14-day) ECG database (14DAYDB). Compute time is reported as the average over all records in the database. SE is sensitivity and PPV is positive predictive value. The grace period is 150 ms.

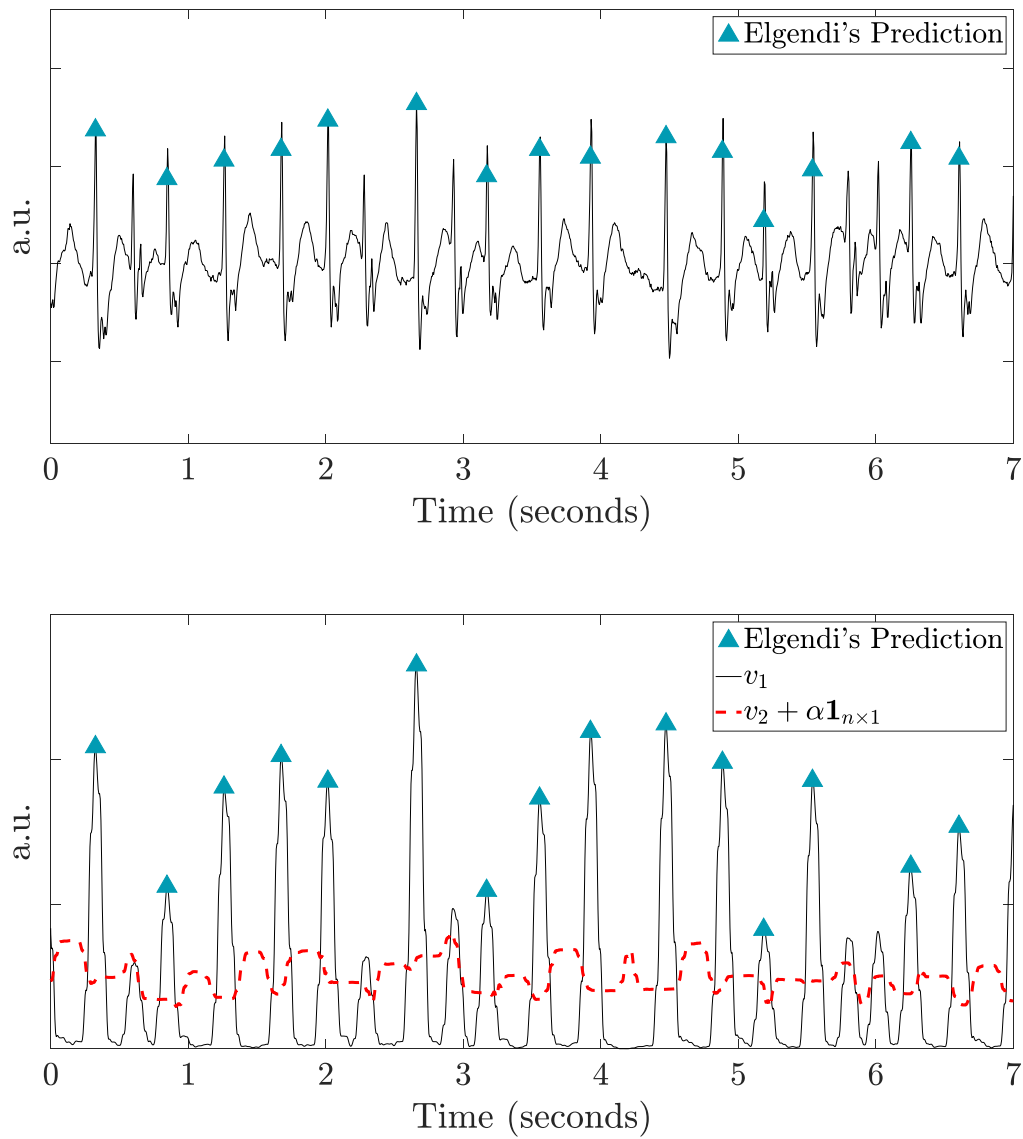

*Figure 1*. Top: a 7-second ECG segment from our new database of long-term and pathological recordings; the QRS complex predictions given by Elgendi's algorithm are shown in blue. The algorithm fails due to the high heart rate. Bottom: the signals $v_1$ (black) and $v_2 + \alpha \mathbf{1}_{n \times 1}$ (red) are shown (see Appendix); false negative beats arise because the red threshold is too large.

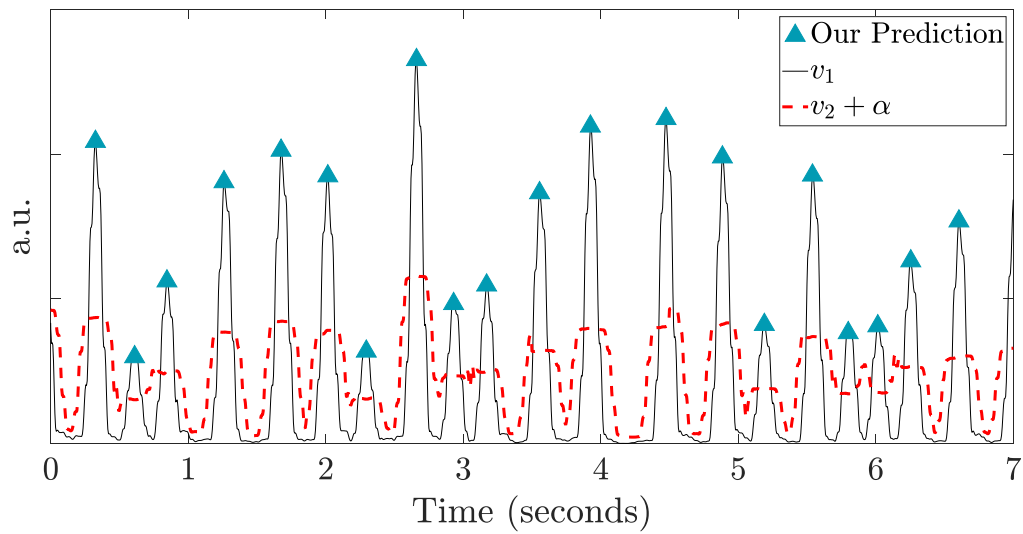

*Figure 2*. We plot the intermediate signals $v_1$ (black) and $v_2 + \alpha$ (red) obtained by running our new QRS complex detection algorithm on the ECG signal shown at the top of Figure 1. The predicted QRS complex locations are shown in blue, and the false negative peaks from Figure 1 are now detected because the red threshold has adapted to the high heart rate.

# Appendix

## **Elgendi's Algorithm (Details)**

Write the raw, single-channel ECG signal as an $n$-dimensional vector $x \in \mathbb{R}^n$, where $n = \lfloor f_s \times T \rfloor$, $f_s$ is the sampling rate of the signal, and $T$ is the duration of the recording in seconds. Write the $i$-th entry of $x$ as $x(i)$. Begin by applying a 3rd order, bi-directional, Butterworth bandpass filter with cutoff frequencies 8 Hz and 20 Hz. Denote the filtered signal as $y \in \mathbb{R}^n$. Form the vector $z \coloneqq y \odot y$ by squaring the entries of $y$. Then, apply two moving-average filters to $z$. Let $W_1$ be the smallest odd integer greater than or equal to $\lfloor 0.097 \times f_s \rfloor$, and let $W_2$ be the smallest odd integer greater than or equal to $\lfloor 0.611 \times f_s \rfloor$. Compute

$$v_1(i) = \frac{1}{W_1}\left[z\left(i - \frac{W_1 - 1}{2}\right) + \cdots + z(i) + \cdots + z\left(i + \frac{W_1 - 1}{2}\right)\right]$$

$$v_2(i) = \frac{1}{W_2}\left[z\left(i - \frac{W_2 - 1}{2}\right) + \cdots + z(i) + \cdots + z\left(i + \frac{W_2 - 1}{2}\right)\right]$$

averaging over the available samples when near the endpoints of the signal $z$. Define

$$\bar{z} = \frac{1}{n}\sum_{i=1}^{n} z(i)$$

and write $\alpha = 0.08 \times \bar{z}$. To detect QRS complexes in the recording $x$, look for sections of the signal $v_1$ which exceed the signal $v_2 + \alpha \mathbf{1}_{n \times 1}$ for a duration of at least $W_1$ consecutive samples. To be specific, if $k_1$ and $k_2$ are two positive integers such that all of the conditions

- $k_2 - k_1 + 1 \geq W_1$
- $v_1(i) > v_2(i) + \alpha$ for all $k_1 \leq i \leq k_2$
- $v_1(k_1 - 1) \leq v_2(k_1 - 1) + \alpha$ or $k_1 = 1$
- $v_1(k_2 + 1) \leq v_2(k_2 + 1) + \alpha$ or $k_2 = n$

hold, then conclude that a QRS complex exists at $x(j)$, where

$$j = \text{argmax}_{k_1 \leq i \leq k_2} v_1(i).$$

Note that since Elgendi does not specify which signal should be used to identify $j$, we took the liberty of selecting $v_1$ (see Discussion).

## **Performance on an Alternate Lead**

To provide further comparisons between our algorithm and Elgendi's algorithm on the conventional ECG databases from Physionet, we report in **Table 2** their performances on the second lead in each record (when available). Note that the FANTASIADB is a single-lead database. Since the STDB has some records that are single-lead, the number of beats for this database is noticeably smaller than the number reported in **Table 1**. Large differences in F1 score when considering the alternate lead (such as on the AFTDB, the MITDB, the NSTDB, the INCARTDB, and the TWADB) could be attributed to the responsiveness of the lead type to ventricular activations, or the choice of lead used by the annotator.

ADAPTIVE QRS DETECTION 26

Table 2: Evaluation of Algorithm Performance on an Alternate Lead

| Database | Number of Beats | Algorithm | SE (%) | PPV (%) | F1 (%) |
|---|---|---|---|---|---|
| AFTDB | 7,590 | This work | 98.99 | 94.81 | 96.85 |
|  |  | Elgendi | 98.51 | 95.79 | 97.13 |
| IAFDB | 7,637 | This work | 86.58 | 82.18 | 84.32 |
|  |  | Elgendi | 85.83 | 82.95 | 84.37 |
| MITDB | 109,494 | This work | 98.81 | 96.28 | 97.53 |
|  |  | Elgendi | 97.92 | 96.76 | 97.34 |
| NSTDB | 25,590 | This work | 98.23 | 89.80 | 93.83 |
|  |  | Elgendi | 98.46 | 89.60 | 93.82 |
| NSRDB | 1,729,629 | This work | 99.98 | 99.21 | 99.59 |
|  |  | Elgendi | 99.98 | 99.24 | 99.61 |
| STDB | 47,345 | This work | 99.88 | 99.17 | 99.53 |
|  |  | Elgendi | 99.89 | 99.21 | 99.55 |
| SVDB | 184,583 | This work | 99.86 | 99.50 | 99.68 |
|  |  | Elgendi | 99.83 | 99.61 | 99.72 |
| QTDB | 86,995 | This work | 99.98 | 98.41 | 99.19 |
|  |  | Elgendi | 99.97 | 98.59 | 99.28 |
| INCARTDB | 175,906 | This work | 99.81 | 99.16 | 99.49 |
|  |  | Elgendi | 99.84 | 99.32 | 99.58 |
| TWADB | 18,991 | This work | 99.87 | 99.52 | 99.70 |
|  |  | Elgendi | 99.83 | 99.41 | 99.62 |

*Note*: The databases are the AF Termination Challenge Database (AFTDB), the Intracardiac Atrial Fibrillation Database (IAFDB), the MIT-BIH Arrhythmia Database (MITDB), the MIT-BIH Noise Stress Test Database (NSTDB), the MIT-BIH Normal Sinus Rhythm Database (NSRDB), the MIT-BIH ST Change Database (STDB), the MIT-BIH Supraventricular Arrhythmia Database (SVDB), the QT Database (QTDB), the St. Petersburg Institute of Cardiological Technics 12-lead Arrhythmia Database (INCARTDB), and the T-wave Alternans Challenge Database (TWADB). SE is sensitivity and PPV is positive predictive value. The grace period is 150 ms.